\documentclass{article}

\usepackage[utf8]{inputenc} 
\usepackage[T1]{fontenc}    
\usepackage{hyperref}       
\usepackage{url}            
\usepackage{booktabs}       
\usepackage{amsfonts}       
\usepackage{nicefrac}       
\usepackage{microtype}      
\usepackage{lipsum}
\usepackage{fancyhdr}       
\usepackage{graphicx}       
\graphicspath{{media/}}     

\usepackage[title]{appendix}
\usepackage{authblk}

\newcommand{\keywords}[1]
{
  \begingroup
  \def\and{, }%
  \small	
  \textbf{Keywords ---} #1
  \endgroup
}

\title{Transforming Redaction: How AI is Revolutionizing Data Protection
}

\author{
  Sida Peng\textsuperscript{1}, Ming-Jen Huang\textsuperscript{2}, Matt Wu\textsuperscript{2}, Jeremy Wei\textsuperscript{2} \\
  \textsuperscript{1}Microsoft, 
  \textsuperscript{2}Foxit Software Inc.
}

\begin{document}
\maketitle

\begin{abstract}
Document redaction is a crucial process in various sectors to safeguard sensitive information from unauthorized access and disclosure. Traditional manual redaction methods, such as those performed using Adobe Acrobat, are labor-intensive, error-prone, and time-consuming. With the burgeoning volume of digital documents, the demand for more efficient and accurate redaction techniques is intensifying.

This study presents the findings from a controlled experiment that compares traditional manual redaction, a redaction tool powered by classical machine learning algorithm, and AI-assisted redaction tools (iDox.ai Redact). The results indicate that iDox.ai Redact significantly outperforms manual methods, achieving higher accuracy and faster completion times. Conversely, the competitor product, classical machine learning algorithm and with necessitates manual intervention for certain sensitive data types, did not exhibit a statistically significant improvement over manual redaction.

These findings suggest that while advanced AI technologies like iDox.ai Redact can substantially enhance data protection practices by reducing human error and improving compliance with data protection regulations, there remains room for improvement in AI tools that do not fully automate the redaction process. Future research should aim to enhance AI capabilities and explore their applicability across various document types and professional settings.
\end{abstract}

\keywords{AI-assisted redaction \and document redaction \and productivity \and sensitive information} 

\section{Introduction}
Document redaction is a critical task across various domains, including legal, medical, and financial sectors. Its primary goal is to conceal or remove sensitive information before document dissemination. Traditional manual redaction methods are labor-intensive, prone to human error, and time-consuming. As the volume of digital documents increases, the need for efficient and accurate redaction methods becomes more pressing.

The significance of document redaction extends beyond mere compliance with privacy regulations. It is an essential practice for maintaining trust and integrity within and outside organizations, particularly vital in sectors such as finance, law, and healthcare, where documents often contain personally identifiable information (PII) and other sensitive data. For instance, in the legal field, inadequate redaction can result in severe legal repercussions, such as hefty fines and loss of client trust. In the medical field, improper redaction can breach patient confidentiality, causing potential harm and legal consequences for healthcare providers. In the financial sector, the failure to protect sensitive financial information can lead to identity theft and financial fraud, undermining the trust between clients and financial institutions. Given the increasing complexity and volume of data, robust redaction practices are more critical than ever.

Manual redaction requires manually scanning documents to identify and obscure sensitive information. This process is not only time-consuming but also susceptible to errors. Human oversight can result in incomplete redaction, where sensitive information remains visible. The repetitive nature of manual redaction tasks can also cause fatigue, increasing the likelihood of errors. Moreover, demographic biases can influence the effectiveness of redaction efforts, with certain groups being more vulnerable to data exposure than others. These challenges underscore the need for more efficient and reliable redaction methods.

The advent of artificial intelligence (AI) has revolutionized various aspects of document processing, including redaction. With advanced machine learning algorithms, AI can swiftly and accurately identify sensitive information within documents, such as personal data, financial details, or classified information, and redact it accordingly. This not only enhances the speed and efficiency of the redaction process but also reduces the risk of human error. AI-driven redaction tools can handle large volumes of documents, making them indispensable for organizations that need to ensure compliance with data protection regulations. Moreover, these tools can be customized to recognize specific types of sensitive information relevant to different industries, further enhancing their utility and precision.

As AI technology continues to evolve, its role in document processing and redaction is likely to expand, offering even more sophisticated and reliable solutions. This study presents the findings from two controlled experiments comparing traditional manual redaction using Adobe Acrobat with AI-assisted redaction using iDox.ai Redact and a competitor product. The results indicate that iDox.ai Redact significantly outperforms manual methods, achieving higher accuracy and faster completion times. Conversely, the competitor product, which necessitates manual intervention for certain sensitive data types, did not exhibit a statistically significant improvement over manual redaction. These findings suggest that while advanced AI technologies like iDox.ai Redact can substantially enhance data protection practices by reducing human error and improving compliance with data protection regulations, there remains room for improvement in AI tools that do not fully automate the redaction process. Future research should aim to enhance AI capabilities and explore their applicability across various document types and professional settings.

\section{Literature Review}

Data leakage poses significant risks across various sectors, particularly in finance, law, and healthcare. In the financial sector, sensitive information such as account details, transaction histories, and personal identification numbers are prime targets for data breaches. Unauthorized access to such information can lead to severe financial losses and damage to the reputation of financial institutions. 
Similarly, the legal sector handles a vast array of confidential information, including client details, case histories, and sensitive legal documents. The exposure of such data can compromise client confidentiality, lead to legal repercussions, and result in substantial financial penalties. Studies such as Mansfield et al. \cite{mansfield2022mask} and Gruschka et al. \cite{gruschka2018privacy} highlight the critical role of accurate redaction in legal contexts to prevent data breaches and protect client privacy. In the healthcare sector, patient records containing medical histories, treatment plans, and personal identifiers are particularly vulnerable. The Health Insurance Portability and Accountability Act (HIPAA) mandates strict regulations for the protection of health information, yet the variability in data formats and the sheer volume of records make compliance challenging. Research by Tucker et al. \cite{Tucker2016} discusses the importance of a thorough process for data anonymization and de-identification, highlighting the necessity of quality control to ensure patient identifiers are properly redacted.

\subsection{AI Enhancements in Document Redaction}
AI technologies have significantly enhanced the capabilities of document redaction by introducing automated and intelligent solutions. Traditional rule-based redaction methods, while useful, are often limited by their inflexibility and inability to adapt to varied data structures and contexts. AI, particularly through machine learning and natural language processing (NLP), has introduced more sophisticated approaches.

For instance, Kraljevic J. \cite{kraljevic2023validating} has demonstrated high accuracy in redacting sensitive information from electronic health records (EHRs) across different hospital systems. The model leverages deep learning to understand context and accurately identify sensitive data, outperforming traditional rule-based methods.
Neerbek J. \cite{neerbek2020sensitive} addresses sensitive information detection in unstructured text documents using deep learning approaches, significantly outperforming previous keyword-based methods.
The C-sanitized model \cite{S_nchez_2015} introduces a privacy model specifically designed for document redaction and sanitization, ensuring a balance between privacy protection and data utility.
The (C, g(C))-sanitization model \cite{S_nchez_2017} further enables an intuitive configuration of the trade-off between the desired level of protection (i.e., controlled information disclosure) and the preservation of the utility of the protected data.
RedactBuster \cite{beltrame2024redactbuster} presents a novel approach for identifying and categorizing redacted entities using AI techniques, significantly improving the accuracy of entity recognition in heavily redacted documents.
The high-throughput machine learning model \cite{zhang2023sensitive} can process vast amounts of data quickly, making them ideal for large datasets commonly found in financial, legal, and medical sectors.
Additionally, the zero-shot learning technique \cite{albanese2023text} has been explored for text sanitization, allowing models to detect and redact sensitive information across various domains without the need for extensive retraining.
Li Y. \cite{li2023privacypreserving} explores methods for enhancing privacy in LLM services through prompt tuning and privatized token reconstruction, ensuring sensitive information protection during training and inference. Kim S. \cite{kim2023propile} introduces a tool for probing privacy leakage in LLMs, highlighting the challenges of detecting and preventing structured and unstructured PII disclosure in LLM outputs.
These studies contribute to a deeper understanding of how AI and advanced privacy models can enhance document redaction processes across various sectors. As AI technology continues to evolve, its role in document processing and redaction is likely to expand, offering more sophisticated and reliable solutions, reducing the risk of human error, and improving compliance with data protection regulations.

\section{Experimental Design}
\label{sec.method}
This section outlines the experimental design used to evaluate the efficacy of AI-assisted document redaction tools. We conducted a controlled experiment comparing traditional manual redaction methods with different AI-assisted techniques. The first arm compares manual redact through Adobe Acrobat with a competitor product powered by classical machine learning algorithm, and the second arm compares manual redact with iDox.ai Redact \footnote{https://www.idox.ai}, an AI-driven tool for identifying sensitive data.

\subsection{Participant Recruitment}
Participants were recruited through Upwork \footnote{https://www.upwork.com}, a well-established freelancing platform. The recruitment process involved advertising the experiments as job postings aimed at freelance document workers. This strategy ensured a diverse pool of participants experienced in document handling. Figure \ref{fig.job.posting} in Appendix \ref{appendix.upwork} illustrates the job posting used, compliant with Upwork’s policies. Upon agreeing to the terms of the experiments, participants were randomly assigned to one of the three groups: the Adobe Acrobat control group, the iDox.ai Redact treatment group, and the group using competitor products.

Figures \ref{fig.adobe.group.email}, \ref{fig.redactable.group.email}, and \ref{fig.idox.group.email} in appendix \ref{appendix.email} display the detailed instructions sent to participants in the control, classical algorithm powered redact tool group, and iDox.ai Redact groups, respectively. Participants were required to watch an instructional video outlining the process of selecting and redacting sensitive data using their assigned tool.

For the classical algorithm powered redact tool group, the instructions included additional guidance on manually selecting sensitive data types. The iDox.ai Redact group received instructions on utilizing the AI-assisted functionality to identify sensitive data automatically. Subsequently, participants received emails containing download links for the documents.

The documents used in these studies contained various categories of sensitive information, necessitating thorough redaction to prevent personal information leaks. Specific redaction targets included personal names, roles or job titles, monetary amounts, number of shares, locations, gender identifiers, addresses, ages, organizations or company names, phone or fax numbers, dates, and dates of birth. Table \ref{table.testfiles} provides a summary of each document's length, word count, and the number of occurrences of sensitive data. This overview helps in understanding the distribution of sensitive information and the overall document size, which are crucial for planning our redaction and data protection efforts.

\begin{table}[ht]
\centering
\begin{tabular}{c|ccc|c}
\hline
& \textbf{Doc 1} & \textbf{Doc 2} & \textbf{Doc 3} & \textbf{Totals} \\ \hline 
\textbf{Pages} & 17 & 12 & 19 & 48 \\ \hline
\textbf{Word Count} & 9,527 & 6,081 & 7,936 & 23,544 \\ \hline
\textbf{Sensitive Data Occurrences} & 110 & 22 & 15 & 147 \\ \hline
\end{tabular}
\caption{Summary of documents and sensitive data occurrences}
\label{table.testfiles}
\end{table}


\subsection{Ethical Considerations}
The study was conducted following ethical guidelines for research involving human participants. All participants provided informed consent and were assured of the confidentiality of their data. The recruitment and experimental procedures were reviewed and approved by an institutional review board (IRB).

\subsection{Limitations}
Despite the robust design, the study has several limitations that should be acknowledged. The reliance on self-reported data for the entry and post-test surveys may introduce biases. Additionally, recruiting participants through a freelancing platform may limit the generalizability of the findings to broader populations. Future research should consider these limitations and aim to address them through alternative recruitment methods and more diverse participant samples.

\section{Results}
\label{sec
}

This section presents the results of the experiments comparing the traditional manual redaction method using Adobe Acrobat with the third party redact tool and the AI powered iDox.ai Redact. The results are summarized in Tables \ref{table.pvalue.redactable} and \ref{table.pvalue.idox}.

\subsection{Manual vs. Classical redact algorithm}
The first arm compares Maunal redact using Adobe Acrobat with classical redact algorithm. The key metrics evaluated were accuracy and time to complete.

\textbf{Accuracy}: We measure the accuracy of the redact task as the percentage of correctly redacted sensitive entries relative to the total identified sensitive entries in the documents. The mean accuracy for the control group (manual) is 91.37\%, while the classical redact algorithm group achieved a mean of 89.48\%. The t-value was 1.29576, and the p-value was 0.198102, indicating that the difference in accuracy between the two groups was not statistically significant.

\textbf{Time to complete}: The time to complete is measured as the total time taken by participants to complete the redaction. The mean time taken for the control group is 19.10 minutes, compared to 17.66 minutes from the classical redact algorithm group. The t-value was 0.7686, and the p-value was 0.443979, suggesting that the difference in time to complete was not statistically significant either.

\begin{table}[ht]
\centering
\begin{tabular}{|c|cccc}
\hline
                               & \textbf{control} & \textbf{treatment} & \textbf{t-value} & \multicolumn{1}{c|}{\textbf{p-value}}  \\ \hline
\textbf{Accuracy}  & 91.37\% & 89.48\% & 1.29576          & \multicolumn{1}{c|}{0.198102}    \\ \hline
\textbf{Time to complete} & 19.10 mins & 17.66 mins  & 0.7686           & \multicolumn{1}{c|}{0.443979}    \\ \hline
\end{tabular}
\caption{P-value of the Accuracy and efficiency of using classical redact algorithm}
\label{table.pvalue.redactable}
\end{table}

\subsection{Manual vs. iDox.ai Redact}

Next, we compare manual redact versus iDox.ai Redact. Again, Accuracy and time to complete were the key metrics evaluated.

\textbf{Accuracy}: The mean accuracy for the control group was 91.37\%, while the treatment group (iDox.ai Redact) achieved a significantly higher mean of 97.10\%. The t-value was -4.29953, and the p-value was 0.00004, indicating a statistically significant improvement in Accuracy with iDox.ai Redact.

\textbf{Time to complete}: The mean time for the control group to complete the redaction task was 19.10 minutes, compared to 15.75 minutes for the treatment group. The t-value was 2.32047, and the p-value was 0.022389, indicating a statistically significant improvement in efficiency with iDox.ai Redact.

\begin{table}[ht]
\centering
\begin{tabular}{|c|cccc}
\hline
                               & \textbf{control} & \textbf{treatment} & \textbf{t-value} & \multicolumn{1}{c|}{\textbf{p-value}}  \\ \hline
\textbf{Accuracy}  &  91.37\% &    97.10\%   & -4.29953         & \multicolumn{1}{c|}{0.00004}    \\ \hline
\textbf{Time to complete} &  19.10 mins  &    15.75 mins   & 2.32047           & \multicolumn{1}{c|}{0.022389}    \\ \hline
\end{tabular}
\caption{P-value of the accuracy and efficiency of using iDox.ai Redact}
\label{table.pvalue.idox}
\end{table}

\subsection{Discussion}
The results of this iDox.ai Redact experiment demonstrate the advantages of using AI-assisted tools for document redaction. The iDox.ai Redact tool not only improved the accuracy of redactions but also significantly reduced the time required to complete the task. These findings underscore the potential for AI-assisted technologies to enhance the efficiency and effectiveness of data protection practices, particularly in environments where large volumes of sensitive information must be processed.
In contrast, the differences between manual and classical redact algorithm were not statistically significant for either metric. This could be attributed to several factors including 
\begin{itemize}
    \item Classical redact algorithm can not fully identify all types of sensitive data. Participants had to manually select certain sensitive data types that were not recognized by the AI. This manual selection process could negate some of the advantages typically offered by AI assistance, such as increased speed and accuracy, thus leading to performance metrics that are similar to those of the control group using Adobe Acrobat.
    \item Participants' familiarity with Adobe Acrobat, a widely used tool, might also play a role. If participants are more accustomed to using Adobe Acrobat, they might perform more efficiently and accurately.
\end{itemize}

\section{Conclusion}
The results of this study demonstrate the superiority of AI-assisted redaction tools over traditional manual methods. The AI-driven iDox.ai Redact tool not only enhanced the accuracy of redactions but also significantly reduced the time required to complete the tasks. This indicates a promising potential for AI technologies to streamline the redaction process, making it more efficient and reliable. The statistical analysis confirms the significant differences in performance, underscoring the practical benefits of adopting AI-assisted solutions in document redaction.

Future research should explore the applicability of AI-assisted redaction across various document types and in different professional settings. Additionally, examining user satisfaction and experience with these tools can provide valuable insights for improving their design and implementation. Addressing the limitations of current AI tools, such as enhancing their capability to identify a broader range of sensitive data types, will be crucial for maximizing their effectiveness.

As AI technology continues to evolve, its integration into document processing and redaction is likely to offer even more sophisticated and reliable solutions. These advancements will ultimately enhance data protection and compliance efforts across multiple sectors, making AI-assisted redaction an increasingly valuable tool in managing sensitive information.

\bibliographystyle{alpha}  
\bibliography{references}

\pagebreak
\begin{appendices}


\section{Upword job posting}
\label{appendix.upwork}

\begin{figure}[ht]
\caption{Upword job posting}
\centering
\includegraphics[width=12cm]{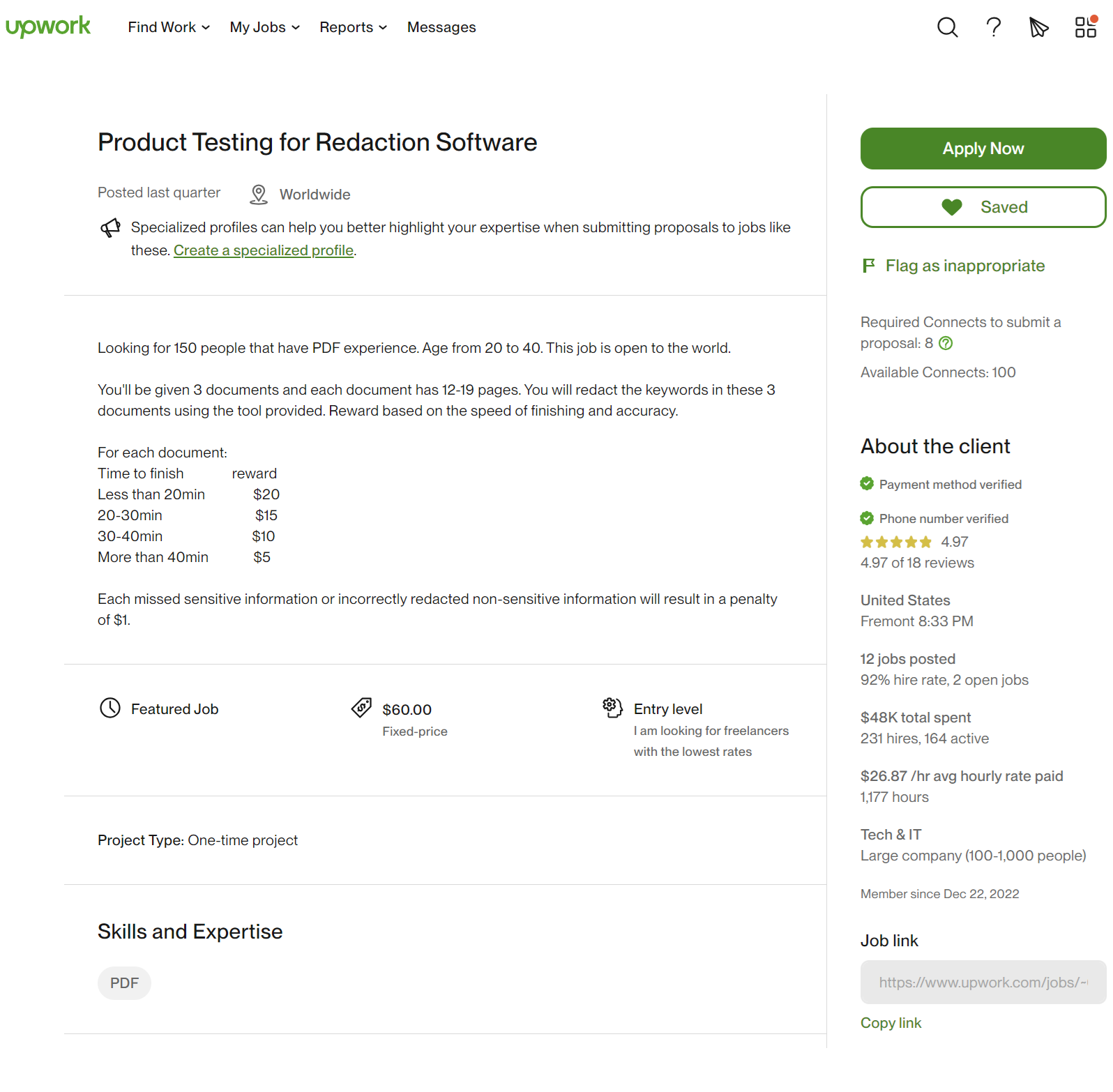}
\label{fig.job.posting}
\end{figure}

\section{Email for the three groups}
\label{appendix.email}
\begin{figure}[ht]
\caption{Email for control group}
\centering
\includegraphics[width=12cm]{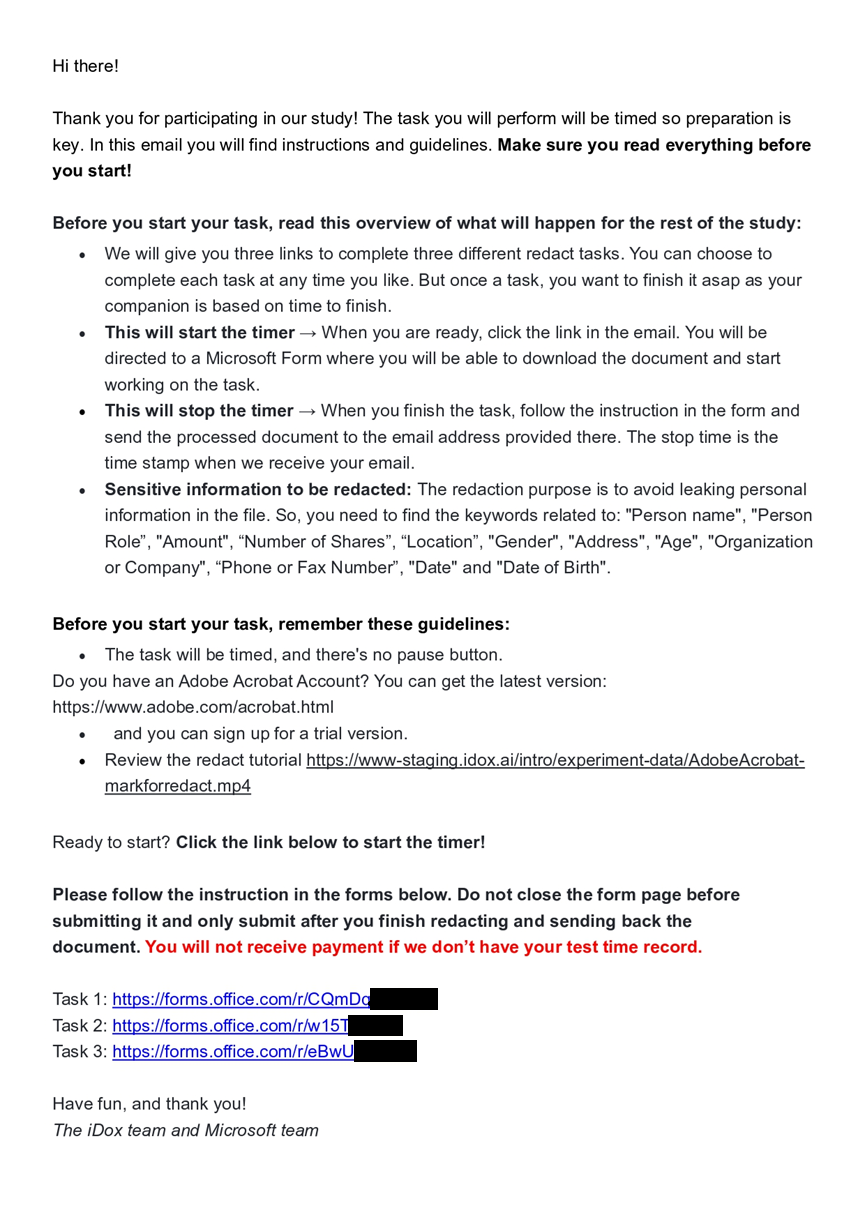}
\label{fig.adobe.group.email}
\end{figure}


\begin{figure}[ht]
\caption{Email for classical redact algorithm group}
\centering
\includegraphics[width=12cm]{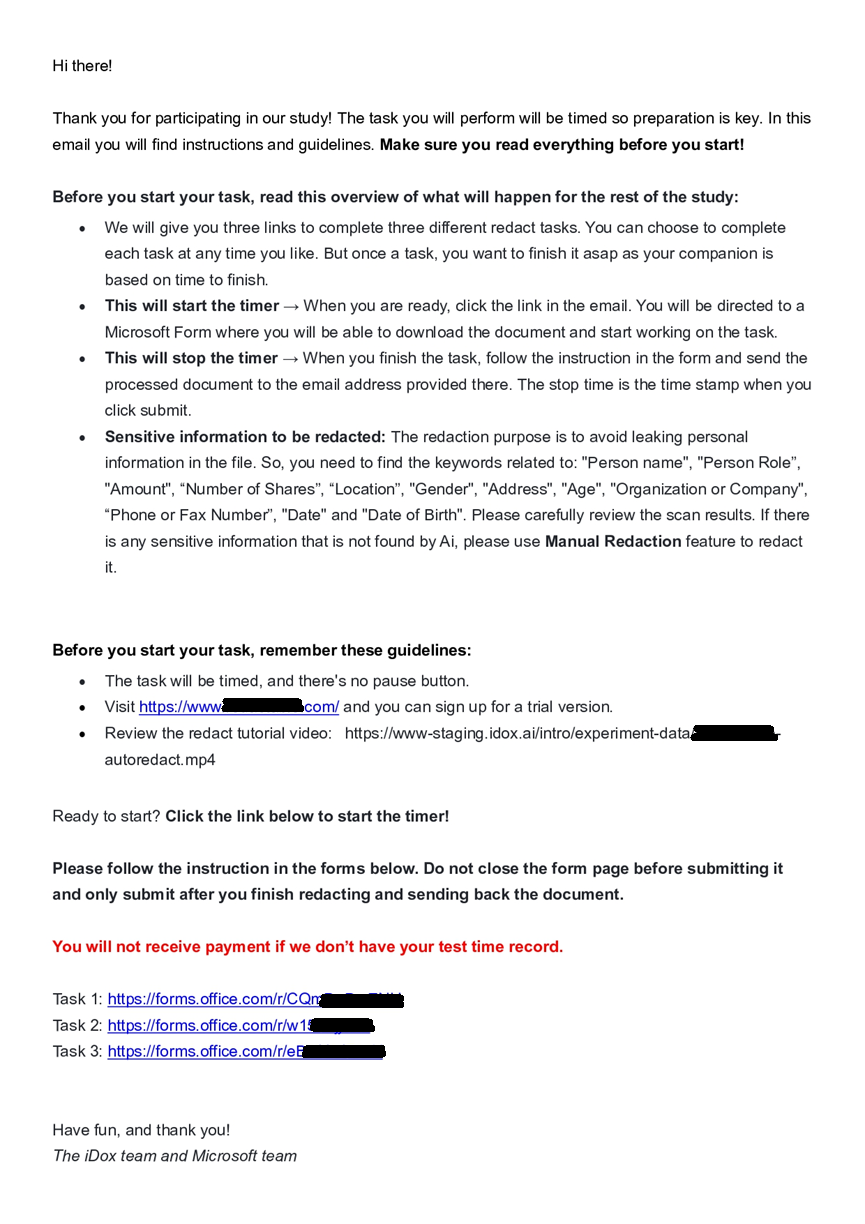}
\label{fig.redactable.group.email}
\end{figure}

\begin{figure}[ht]
\caption{Email for iDox.ai Redact group}
\centering
\includegraphics[width=12cm]{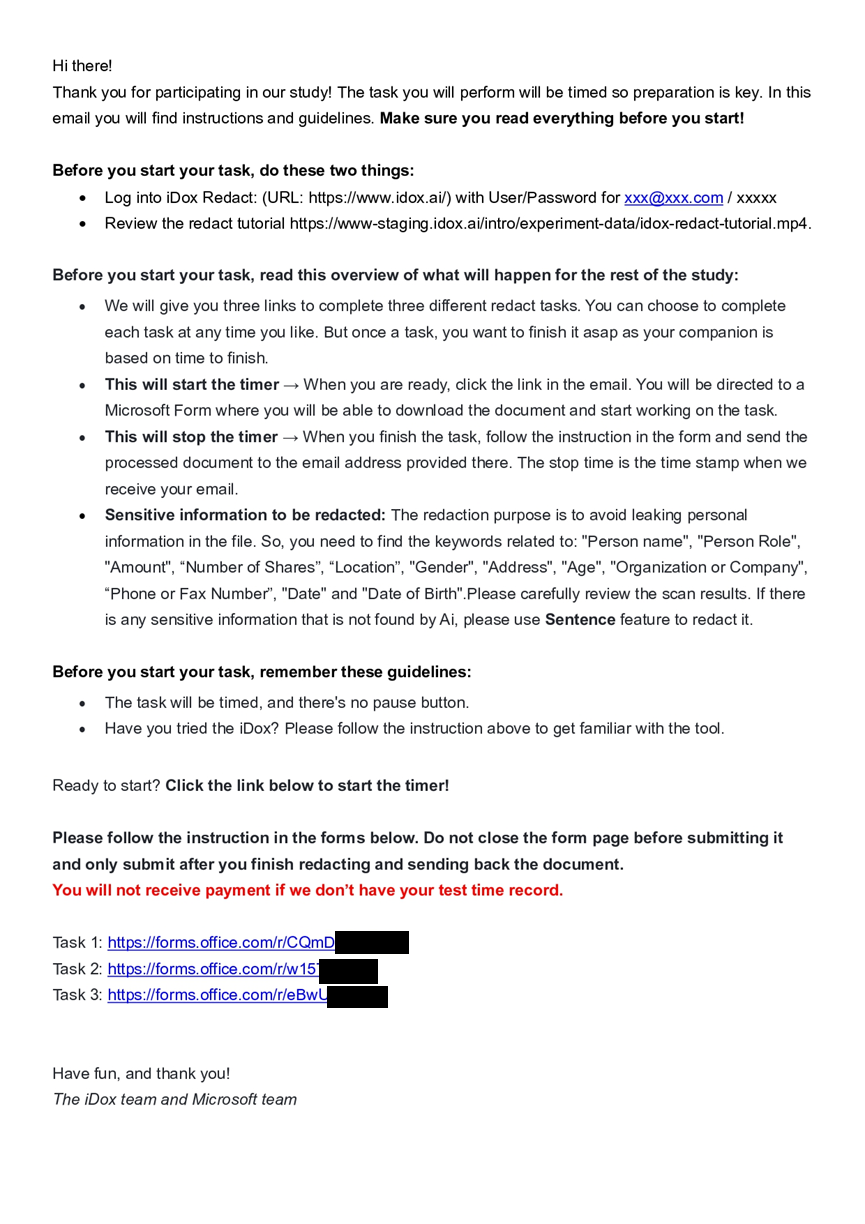}
\label{fig.idox.group.email}
\end{figure}
\end{appendices}

\end{document}